%% file: paper.tex
\documentclass{sig-alternate}

\input{macros}

\def\sharedaffiliation{%
\end{tabular}\\
\begin{tabular}{c}}

\begin{document}

\title{Notable Characteristics Search through Knowledge Graphs}

\numberofauthors{5}

\author{
Davide Mottin\\
\email{{\small davide.mottin@hpi.de}}
\and
Bastian Grasnick\\
\email{{\small bastian.grasnick@student.hpi.de}}
\and
Axel Kroschk\\
\email{{\small axel.kroschk@student.hpi.de}}
\and 
Patrick Siegler\\
\email{{\small patrick.siegler@student.hpi.de}}
\and
Emmanuel M\"uller\\
\email{{\small emmanuel.mueller@hpi.de}}
\sharedaffiliation
\affaddr{Hasso Plattner Institute}
}

\maketitle

\begin{abstract}
\input{sections/abstract}

\end{abstract}

\input{sections/introduction}

\input{sections/problem}

\input{sections/solution}
\input{sections/experiments}
\input{sections/related}

\input{sections/conclusions}

\clearpage

\balance

\bibliographystyle{abbrv}

\end{document}

%% file: macros.tex
\usepackage[show]{chato-notes}
\usepackage{url}
\usepackage[pdfpagelabels=false]{hyperref}
\usepackage{graphicx}
\usepackage{multirow}
\usepackage{amssymb}
\usepackage{amsmath}
\usepackage{centernot}

\usepackage{amsthm}
\usepackage{algorithm}
\usepackage{algorithmicx}
\usepackage[noend]{algpseudocode}
\usepackage{enumerate}
\usepackage{colortbl}
\usepackage{comment}
\usepackage{balance}
\usepackage[tikz]{bclogo}
\usepackage[normalem]{ulem}
\usepackage{tikz}
\usepackage{nccmath}
\usepackage{enumitem}
\usepackage{balance}
\usepackage[nocompress]{cite}
\usepackage{subcaption}

\usetikzlibrary{decorations.markings}

\hypersetup{
    unicode=false,          
    pdftoolbar=true,        
    pdfmenubar=true,        
    pdffitwindow=false,     
    pdfstartview={FitH},    
    pdftitle={My title},    
    pdfauthor={Author},     
    pdfsubject={Subject},   
    pdfcreator={Creator},   
    pdfproducer={Producer}, 
    pageanchor=false,
    pdfnewwindow=true,      
    colorlinks=true,       
    linkcolor=black,          
    citecolor=black,        
    filecolor=black,      
    urlcolor=black,           
}

\newcommand*{\graphquad}[4]{\langle #1, #2, #3, #4 \rangle}
\newcommand*{\sequence}[2]{\langle #1_1, ..., #1_#2 \rangle}

\newcommand*{\labelset}{\mathcal{L}}

\newcommand*{\vertices}{\mathcal{V}}

\newcommand*{\edges}{\mathcal{E}}

\newcommand*{\ourprob}{notable characteristics search}
\newcommand*{\Ourprob}{Notable characteristics search}

\newcommand*{\nodelabels}{\mathcal{A}}
\newcommand*{\edgelabels}{\mathcal{L}}

\newcommand*{\seed}{Q}
\newcommand*{\context}{C}
\newcommand*{\simfun}{\sigma}
\newcommand*{\discfun}{\delta}

\newcommand*{\Rpluszero}{\mathbb{R}^+_0}
\newcommand*{\instdist}[1]{Inst_{#1}}
\newcommand*{\carddist}[1]{Card_{#1}}
\newcommand*{\instdiscfun}{\discfun_{Inst}}
\newcommand*{\carddiscfun}{\discfun_{Card}}
\newcommand*{\mtfun}{\mathrm{MT}}

\newcommand*{\actors}{actors}
\newcommand*{\politicians}{politicians}
\newcommand*{\moviepeople}{movie contributors}

\newcommand*{\ouralgorithm}{\textsc{ContextRW}}
\newcommand*{\findnotable}{\textsc{FindNC}}
\newcommand*{\rw}{\textsc{RandomWalk}}
\newcommand*{\rwtest}{\textsc{RWMult}}

\newcommand*{\yago}{\textsf{YAGO}}
\newcommand*{\lmdb}{\textsf{LinkedMDB}}

\newcommand{\spara}[1]{\smallskip\noindent{\bf #1}}
\newcommand{\mpara}[1]{\medskip\noindent{\bf #1}}

\newtheorem{problem}{Problem}

\newtheorem{definition}{Definition}

\newtheorem*{example*}{Example}
\newtheorem*{problem*}{Problem}
\newtheorem*{theorem*}{Theorem}
\newtheorem*{definition*}{Definition}
\newtheorem*{lemma*}{Lemma}
\newtheorem*{property*}{Property}

%% file: sections/abstract.tex

Search engines employ complex data structures to assist the user in the search process. 
Among these structures, knowledge graphs are vastly used for various search tasks. 
Given a knowledge graph that represents entities and relationships among them, one aims at complementing the search  with intuitive but effective mechanisms.
In particular, we focus on the comparison of two or more entities and the detection of unexpected properties, called \emph{notable characteristics}.
These notable characteristics find large applicability in many domains since they provide non-trivial insights of the entities into consideration in an intuitive and domain-independent fashion. 
To this end, we propose a novel formulation of the problem of searching and retrieving notable characteristics given an initial set of query nodes. 
While the traditional comparison of nodes by means of node similarity provides only a score with no explanation, we go one step further.
We propose a solid probabilistic approach that first retrieves nodes that are similar to the query nodes provided by the user, and then exploits distributional properties to understand whether a particular attribute is interesting or not. 
We experimentally evaluate the effectiveness of our approach and show that we are able to discover notable characteristics that are indeed interesting and relevant for the user.

%% file: sections/introduction.tex
\section{Introduction}
\label{sec:intro}

Search engines have greatly evolved from simple indexes of pages to complex systems that are able to predict user intention, show personalized content, answer queries on a large variety of data sources. 
One way to improve the search quality is by using a knowledge graph representation of data including relationships among entities. 
A knowledge graph represents entities (e.g., Barack Obama, USA) as nodes and relationships between them (e.g., leaderOf) as edges in a graph.
With this representation knowledge graphs empower search capabilities by exploiting the relations among entities~\cite{lissandrini2015unleashing}. They have been successfully employed for text understanding~\cite{hua2015short}, keyword search expansion~\cite{bordino2013machu_picchu}, and semantic reasoning~\cite{wang2004ontology}.

\begin{figure}[t!]
    \centering
    \includegraphics[width=0.47\textwidth]{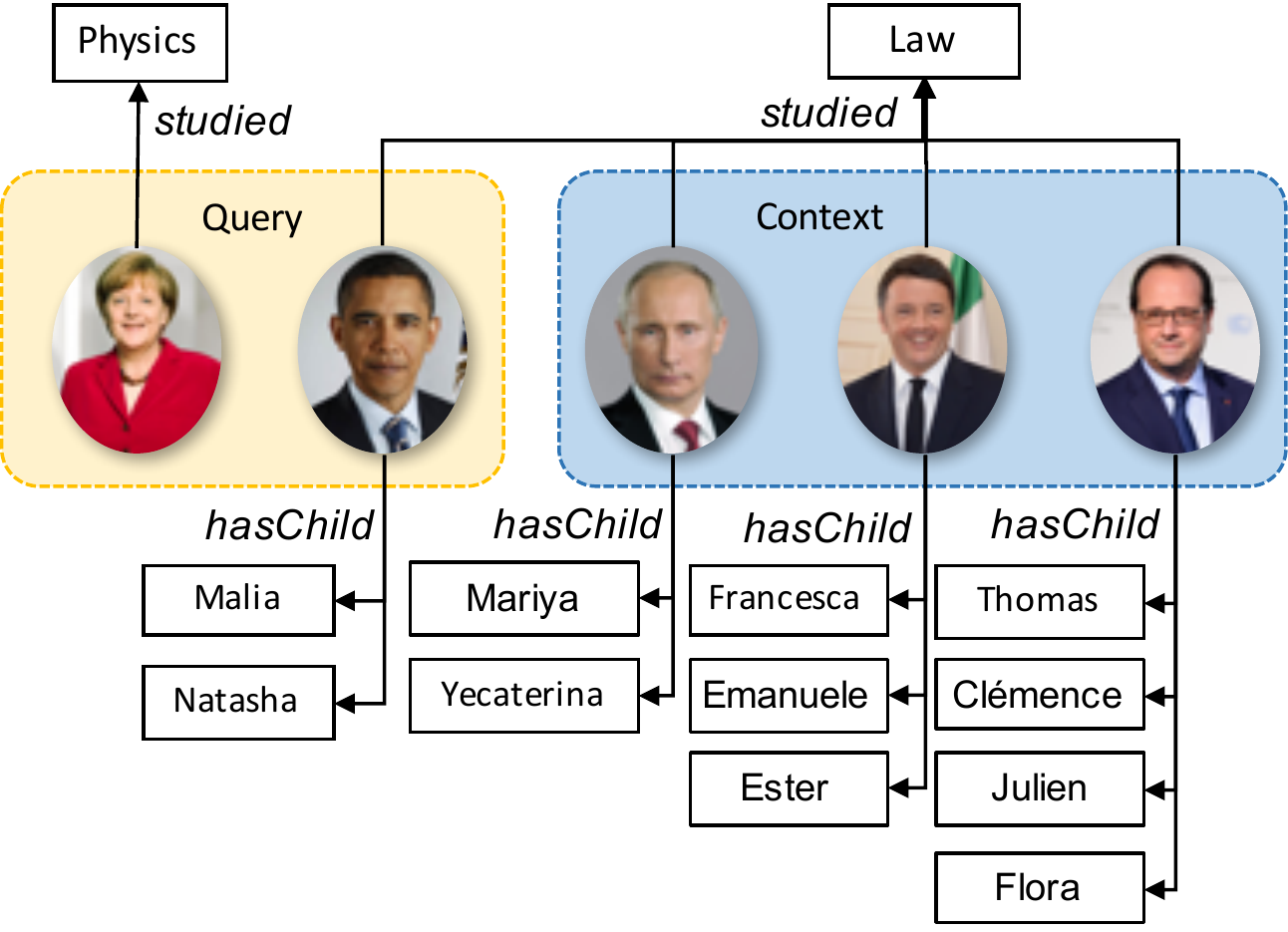}
    \caption{An example knowledge graph, the query (Merkel and Obama), and the discovered context nodes (Putin, Renzi, and Hollande). The fact that Angela Merkel does not have a child is a notable characteristic.}
    \label{fig:motivating_example}
\end{figure}

The great expressiveness of knowledge graphs can complement the search with more flexible search paradigms. 
Assume for instance a scholar who requires to know some non-trivial facts about Angela Merkel and Barack Obama with respect to other country leaders. 
It would be interesting to discover for instance that Angela Merkel has a PhD as opposed to most of the other leaders, and that she has no children. 
We call this fact a \emph{notable characteristic}, to remark the unexpected and non-trivial aspect of the discovery. 
In this work we propose a novel type of search called \emph{notable characteristics search} that allows the retrieval of such facts from a set of input query entities.
Discovering notable characteristics also constitutes a ground for targeted analyses of products in electronic commerce or microorganisms in biological networks. 
Imagine a user compares two cameras and wants to know what are the special features of these two with respect to all the others. 
In general it can be used for large graphs that are hard to explore manually and need assistance in the discovery of notable characteristics, as a mean of comparison between entities.
As a consequence, in all the cases in which a knowledge graph is available, the discovery of notable characteristics becomes an expressive and powerful search type for any user, from experts and practitioners to novice users. 
Moreover, the use of graphs allows for the definition of domain independent graph techniques that can easily adapt to different networks.

In our setting, we assume the user provides a set of \emph{query} nodes to be compared and the algorithm finds a set of notable characteristics of these nodes. 
We note that having nodes as input is not a restriction to the generality of the method since there exists a number of techniques that correctly map keywords to nodes in any knowledge graph~\cite{kasneci2009ming,pound2012interpreting}. 
Given a node, a property is a relationship with other nodes (e.g., leaderOf). 
A characteristic or property is notable, if it deviates from what one would expect for the nodes into consideration.
To the best of our knowledge, this is the first study of automatic discovery of notable characteristics (or properties). 

The discovery of notable characteristics entails two challenges. 
First, given the set of query nodes we need to compare them to only those nodes that are similar to some extent. 
Second, we need to select only those properties that are significantly different from the one expressed in the query.
Please note that tackling the first challenge is very important, as comparison of the query nodes has to be performed with a set of similar nodes, which we call the context of the query.
Consider the na\"ive approach that returns notable characteristics simply by comparing the query nodes and assume that the user provides ``Angela Merkel'' and ``Dilma Roussef'' as query. 
This is a counter example for the na\"ive direct comparison, as it will not return the gender as a notable characteristic. 
Both query nodes are female, however only in comparison with other presidents this becomes an interesting fact. 
On the other extreme, selecting all the nodes in the graph as context will mislead the analysis towards non-relevant nodes.
Take our example of ``Angela Merkel'' and ``Barack Obama''. 
A na\"ive selection of all humans will not work as context, since the gender is equally distributed in the world as well as in the input nodes, the fact ``Angela Merkel is a woman'' is not notable.
 
Hence, it is crucial to provide a thorough context selection to prevent the above cases. 
Therefore, we introduce the discovery of \emph{context nodes}, i.e., nodes similar to the query nodes. 
An example of the proposed approach is depicted in in Figure~\ref{fig:motivating_example}.
To this end, we devise a method that exploits metapaths~\cite{sun2011pathsim} and random walks for context discovery.
We also propose a generic framework that efficiently discover notable characteristics through a novel probabilistic approach based on distribution comparison.

Our contributions are summarized as follows: 
\begin{itemize}
\setlength\itemsep{0em}
	\item We formalize the problem of notable characteristics search given a set of query nodes as input.
	\item We show how to effectively compute metapaths to find the context nodes in knowledge graphs. 
	\item We introduce a probabilistic approach to discover notable characteristics given a query node set.
	\item We experimentally evaluate our context selection approach through a user study, and show evidence of our discovered notable characteristics and the real time performance of the proposed algorithms.
\end{itemize}

The paper is structured as follows. 
Section~\ref{sec:problem} introduces the problem of notable characteristics search given a set of nodes. 
In Section~\ref{sec:solution} we present our solution based on random walk constrained with metapath discovery and the probabilistic framework to identify notable characteristics. 
The solution is empirically evaluated in Section~\ref{sec:experiments}. 
We present the related work in Section~\ref{sec:related}, and finally conclude in Section~\ref{sec:conclusions} with remarks and future work.

%% file: sections/problem.tex
\section{Problem statement}
\label{sec:problem}

In this section, we introduce the problem of \emph{\ourprob} in a knowledge graph given a set of input \emph{query nodes}.
A knowledge graph is a directed graph in which nodes and edges have labels or types. 
They are also known as information networks~\cite{meng2015discovering,lee2012pathrank} or simply labeled graphs.
We are given a set $\nodelabels$ of node labels and a set $\edgelabels$ of edge labels.
The term label and type are used interchangeably.

\begin{definition}[Knowledge graph]
\label{def:knowledge-graph}
A knowledge graph is a quadruple $G:\graphquad{\vertices}{\edges}{\phi}{\psi}$, where $\vertices$ is a set of nodes, $\edges \subseteq \vertices \times \vertices$ is a set of edges, $\phi: \vertices \mapsto \nodelabels$, $\psi: \edges \mapsto \edgelabels$ are node and edge labeling functions, respectively.  
\end{definition}

For simplicity, we assume that everything is modeled as relationships and nodes. 
This is the case for attributes such as birth date: we assume that the date itself is a node connected with a ``birthdate'' relationship.
Additionally, we assume that for every edge $e \in \edges$ with type $\psi(e) = l$ exists a reverse edge $e^{-1}$ with $\psi(e^{-1}) = l^{-1}$ to model cases such as ``presidentOf'' and ``hasPresident''.
The above assumptions do not change the nature and the generality of the methods but simplify the notation and the analysis. 

Recall that we are interested in discovering \emph{notable characteristics} of the entities mentioned in a set of input query nodes in relation to their similars.
This intuitive definition entails two questions: (1) what is the set of similars? (2) what are the notable characteristics? 

The set of input nodes is referred to as \emph{query set} (query in short). 
Formally, given a knowledge graph $G: \graphquad{\vertices}{\edges}{\phi}{\psi}$ the query is any set $\seed \subseteq \vertices$.
The query set is manually provided by the user and therefore considered reasonably small (i.e., $\leq10$ elements).
The first question concerns the definition of a set of similars referred in this work as \emph{context nodes}. 
We assume the existence of a \emph{similarity function} $\simfun : \vertices \times 2^\vertices \mapsto \mathbb{R}$ that assigns a high score to nodes that are similar to those in the query set and low otherwise.
Then, the context are the top-$k$ most similar nodes.

\begin{definition}[Context set]
\label{def:context}
Given a knowledge graph $G : \graphquad{\vertices}{\edges}{\phi}{\psi}$, a query set $\seed \subseteq \vertices$, a similarity function $\simfun : \vertices \times 2^\vertices \mapsto \mathbb{R}$, and a parameter $k$, the \emph{context set} (or simply context) is a set $\context \subseteq \vertices$ such that $\seed \cap \context = \emptyset, |\context| = k$, and for each $n_c \in \context \wedge n \in \vertices \setminus (\seed \cup \context), \simfun(n,\seed) \le \simfun(n_c,\seed)$.
\end{definition}

The second question concerns the notable characteristics. 
The characteristics are attributes or relationships of a specific node since they implicitly represent a signature of the node itself. 
As before, we assume the existence of a generic \emph{discrimination function}, whose role is to return a score whether a specific characteristic is discriminative or unexpected comparing two set of nodes. 
Formally, in the knowledge graph $G$, a discrimination function $\discfun: \edgelabels \times 2^\vertices \times 2^\vertices \mapsto \Rpluszero$ assigns a discrimination value or 0 if the value is not discriminative. 
We are now ready to define a notable characteristic. 

\begin{definition}[Notable characteristic]
\label{def:notable}
Given a knowledge graph $G : \graphquad{\vertices}{\edges}{\phi}{\psi}$, a query $\seed \subseteq \vertices$, a context $\context \subseteq \vertices$, and a discrimination function $\discfun: \edgelabels \times 2^\vertices \times 2^\vertices \mapsto \Rpluszero$ a \emph{notable characteristic} is a relationship $l \in \edgelabels|_{\seed \cup \context}$ such that $\discfun(l,\seed,\context) \neq 0$.
\end{definition}

The notation $\edgelabels|_{\seed \cup \context}$ denotes the set of edge labels restricted to those that are found in the edges directly connected to $\seed \cup \context$, i.e., $\edgelabels|_{\seed \cup \context} = \{l~|~\exists x \in \seed \cup \context, y \in \vertices  \text{ s.t. } (x,y) \in \edges \wedge \psi(x,y) = l \}$. 

The general problem we aim to solve is efficiently returning the notable characteristics, given a query, a similarity function and a discrimination function. 

\begin{problem}[\Ourprob]
Given a knowledge graph $G:\graphquad{\vertices}{\edges}{\phi}{\psi}$, a query $\seed \subseteq V$, a similarity function $\simfun : \vertices \times 2^\vertices \to \mathbb{R}$ and a discrimination function $\discfun: \edgelabels \times 2^\vertices \times 2^\vertices \mapsto \Rpluszero$, find the set of notable characteristics. 
\end{problem}

The problem entails the definition of appropriate $\simfun$ (similarity) and $\discfun$ (discrimination) functions that retrieve and compare nodes in the knowledge graph. 
In Section~\ref{sec:solution} we provide an elegant instance by means of a probabilistic framework that is able to discover meaningful results. 
We also provide the motivation of our choices by considering several variants of the above functions.

%% file: sections/solution.tex
\section{Notable characteristics search}
\label{sec:solution}

In this section we describe methods to automatically discover notable characteristics given a set of query nodes. 
Recall that the problem requires the definition of a similarity function~$\simfun$ and a discrimination function~$\discfun$. 
We model the discrimination function in probabilistic terms, in order to better deal with noisy settings and uncertainty. 
Therefore, we assume that a characteristic is interesting if its distribution in the query nodes deviates from the one in the context set. 
In other words, the context represents the expected behavior of the population while the query is the hypothesis to be tested.

Section~\ref{ssec:context} shows how to effectively find the context nodes, while Section~\ref{ssec:distributions} describes the comparison of distributions to effectively discover notable characteristics.

\input{sections/context}

\input{sections/distribution}

%% file: sections/context.tex
\subsection{Finding the context}
\label{ssec:context}

Given the query $\seed$, we define a similarity function $\simfun$ to retrieve a set of context nodes. 
Although many notions of similarity functions have been developed, such as structural equivalence~\cite{lorrain1971structural} and SimRank~\cite{jeh2002simrank}, none seems suitable to our case. 
Existing similarity measures are either based on restricted neighborhoods of the nodes~\cite{lorrain1971structural}, or they disregard edge and node labels~\cite{jeh2002simrank}.
We devise an algorithm that takes into account edge labels and combines the advantages of random walk and metapath approaches. 

In the traditional random walk model, a random walker chooses one of the outgoing edges from a node with uniform probability. 
Instead of uniform probability, we favor choices which are more informative in terms of edge label frequency: 
the lower the frequency the more informative the label. 
This intuition is supported by information theoretic notions, such as tf-idf and has been successfully used in graphs as well~\cite{mottin2014exemplar}. 
As a shorthand notation, we define $\edges_l$ as the set of edges having label $l \in \labelset$, i.e., $\edges_l = \{(i,j) \in \edges | i,j \in \vertices, \psi(i,j) = l\}$.
The frequency of a label $l$ is the fraction of $l$-labeled edges with respect to the total number of edges. 
We then define the weighted adjacency matrix as a $|\vertices|\times|\vertices|$ square matrix, where the value $A_{ij}$ between node $i$ and $j$ is defined as

\begin{equation}
\label{eq:adjmatrix}
 A_{ij} = \left\{ 
\begin{array}{ll}
1 - |\edges_l|/|\edges| & \text{if }(i,j) \in E\\
0 & \text{otherwise}
\end{array}
\right. 
\end{equation}

The Personalized PageRank is defined as the vector

\begin{equation}
\label{eq:pagerank}
\mathbf{p} = c\tilde{A}\mathbf{p} + (1 - c)\mathbf{v},
\end{equation}

where $\tilde{A}_{ij} = A_{ji}/\sum_{k}A_{jk}$, $c$ is the damping factor, and $\mathbf{v}$ is vector called personalization vector. 
In our experiments the damping factor is $0.8$, in line with previous works.
We compute the PageRank starting from each node in the query to retrieve the $k$ nodes with the highest score. 
This is done by setting $\mathbf{v}_n = 1$ for each $n \in \seed$, individually.
We refer to this baseline as \rw. 

However, the \rw\ baseline disregards common connections between the query nodes.
This is an important information, since the frequency-based approach does not consider the user's similarity notion implicitly contained in the query.
To this end, we adopt the notion of metapath from~\cite{sun2011pathsim,lee2012pathrank} which generalizes the concept of path.
A metapath for a path $\sequence{n}{t}, n_i \in \vertices, 1 \leq i \leq t$ is a sequence $\langle \phi(n_1),\psi(n_1, n_2), ... ,\psi(n_{t-1}, n_t), \phi(n_t)\rangle$ that alternates node and edge labels along the path.

We mine metapaths as follows. 
We sample a node in $\vertices \setminus \seed$ with uniform probability and run a random walk until a query node is reached. 
The sequence of edge labels $m$ encountered during the random walk is added to the set of metapaths $M$ along with the number of times $c(m)$ the same metapath has been found so far. 
It has been proved that random walks are effective in mining metapaths~\cite{lee2015pathmining}.

Once the metapaths are retrieved, we compute a score for each node based on the probability that some metapath starting from a query node ends in this node. 
Given the set of metapaths $M$, we denote as $\{n \overset{m}{\rightsquigarrow} n'\}$ the set of paths from node $n$ to $n'$ matching metapath $m \in M$.
The score of a node $n' \in \vertices \setminus \seed$ with respect to any query node $n \in \seed$ is
\[
\simfun(n',\seed) = \sum_{m \in M, n \in \seed}\frac{|\{n \overset{m}{\rightsquigarrow} n'\}|}{|\{n \overset{m}{\rightsquigarrow} n'' | n'' \in \vertices \setminus \seed\}|} \mathrm{Pr}(m)
\]
$\mathrm{Pr}(m)$ is the probability of choosing metapath $m$, which is the relative count computed previously divided by the sum of the counts of all metapaths $M$, i.e., $c(m)/\sum_{m\in M}c(m)$. 
Intuitively, $\simfun$ gives a higher score to nodes that are reachable through frequent metapaths connecting the query nodes or connected through many of these metapaths. 
Hence, nodes that are reached from infrequent metapaths will have a low score. 
Once we have computed the score for each node we return the $k$ nodes with the highest score as our context.

%% file: sections/distribution.tex

\subsection{Comparing the distributions}
\label{ssec:distributions}

We revise the definition of notable characteristics in probabilistic terms. 
Assume we have computed the distribution of values for each characteristic (i.e., edge label) for both query and context nodes found with the method in Section~\ref{ssec:context}. 
Intuitively, for each characteristic, the distribution of the context represents the expected, or normal behavior. 
Therefore, the query set becomes the hypothesis to be evaluated against the ``true'' distribution of the context. 

Formally, for each characteristic $l \in \edgelabels$, we consider two distributions in order to evaluate its notability. 
The first represents the frequency of the node labels (e.g., California) connected to a specific edge label (e.g., bornIn). 
This expresses information about the actual values in the nodes and can be used to identify cases where different attribute values are relevant. 
For instance, most people in the query in Figure~1 are half American and half European, while those in the context are all Europeans.
We refer to these distributions as \emph{instance distributions}.
\begin{equation*}
	\instdist{q}(l, \context, \seed) = (x_1, x_2, ..., x_t)
\end{equation*}
\begin{equation*}
	\instdist{c}(l, \context) = (y_1, y_2, ..., y_t)
\end{equation*}
where $x_i$ and $y_i$ are the number of occurrences of node $i$ at the end of an edge labeled $l$ from a node in $\seed$ and $\context$, respectively.
In the example in Figure~\ref{fig:motivating_example}, $\instdist{q}(studied, \context, \seed) = (1,1)$, $\instdist{c}(studied, \context) = (0,3)$, where the first position in the vector indicates Physics studies and the second Law. 
Note that both vectors have the same size, so $x_i$ is zero if $i$ appears only in the context.

The second distribution computes aggregates over the number of occurrences of a specific edge label in the context.
This expresses information about the existence and cardinality of an attribute and can be used to identify cases where attribute cardinality is relevant.
For instance ``Angela Merkel'' in the query in Figure~\ref{fig:motivating_example} has no child, while in the context all other leaders have at least one.
Such cases cannot evidently be modeled as instance distributions that take into account distinct values (e.g., the child name).
We refer to these distributions as \emph{cardinality distributions}.
\begin{equation*}
	\carddist{q}(l, \context, \seed) = (x_1, x_2, ..., x_t)
\end{equation*}
\begin{equation*}
	\carddist{c}(l, \context) = (y_1, y_2, ..., y_t)
\end{equation*}
where $x_i$ and $y_i$ are the number of times a node in $\seed$ and $\context$ respectively has $i$ edges labeled $l$.

Both distributions can be built by iterating through the nodes in each set and counting the respective occurrences.
For a given $l \in \edgelabels$, this results in two scores $\instdiscfun$ and $\carddiscfun$.
The final score $\discfun$ is the maximum score between $\instdiscfun$ and $\carddiscfun$. 
\begin{equation}
\label{eq:discfun}
	\discfun(l, \context, \seed) = \max(\instdiscfun(l, \context, \seed), \carddiscfun(l, \context, \seed))
\end{equation}

Many measures have been proposed in statistics to compare two distributions.  
However, most of them draw specific assumptions, such as a minimum number of samples or non-zero probabilities, that are not fulfilled in our case.
In particular, $\instdist{}$ and $\carddist{}$ have no natural ordering and no distance-function between the values. 
Additionally, the context has a much large variety of node labels than the query.  
This leads to many zero values in the query-distribution. 
Therefore, the commonly used Kullback-Leibler divergence (KL)~\cite{kullback1951information} cannot be used. 
Classical statistical tests, such as the z-test and the $\chi^2$ test require either a Gaussian distribution or a minimum size of the sample. 
On the other hand, the Earth Mover's Distance (EMD) requires the definition of distance between values, which is not defined for $\instdist{}$.

In conclusion, we resorted to a more natural multinomial test that better expresses the relationship between our distributions.
The multinomial test assumes that a set of observations is drawn from a known multinomial distribution. 
Therefore, assuming the context to be multinomial distributed the observations are the values found in the query. 
If the values observed in the query are drawn from the multinomial, than the hypothesis cannot be rejected and the characteristic is marked as non-notable. 
On the other hand, if the test succeeds, then the two distributions are significantly different and the characteristic is notable. 

Assume we have a random variable $X_{N,\pi} \sim Mult(N, \pi)$, with parameters $N$ and $\pi$. 
We normalize $\instdist{c}$ and $\carddist{c}$ to express a probability distribution $\pi =  normalize(y) = (\pi_1, \pi_2, ..., \pi_k)$.
For a given outcome $x = (x_1, x_2, ..., x_k)$, the probability, under the hypothesis of equality between context and sample, is
\begin{equation*}
	\mathrm{Pr(X_{N, \pi} = x)} = N! \prod_{i=1}^k \frac{\pi_i^{x_i}}{x_i!},
\end{equation*}
where $N = \sum x_i$.
In an exact multinomial test, the significance probability is
\begin{equation*}
	\mathrm{Pr}_s(X_{N,\pi} = x) = \sum_{y: \mathrm{Pr}(X_{N,\pi} = y) ~\le~ \mathrm{Pr}(X_{N,\pi} = x)} \mathrm{Pr}(X_{N,\pi} = y)
\end{equation*}

$\mathrm{Pr_s}(\pi,x)$ is the probability of $x$ or any equally or less likely outcome being drawn from the probability distribution\footnote{In case of large $N$, the exact test is impractical, a Montecarlo sampling to approximate the final result is performed.}.
A difference in distributions is considered significant if the hypothesis is rejected with probability $p > 0.95$.

\begin{equation*}
	\mtfun(\pi, x) = \begin{cases}
		1 - \mathrm{Pr_s}(X_{N, \pi} = x) &\mbox{if } \mathrm{Pr_s}(...) \leq 0.05 \\
		0 &\mbox{otherwise}
	\end{cases}
\end{equation*}

Finally, $\discfun$ can be defined as

\begin{equation*}
	\instdiscfun(l, \context, \seed) = \mtfun(norm(\instdist{c}(l, \context)), \instdist{q}(l, \context, \seed))
\end{equation*}
\begin{equation*}
	\carddiscfun(l, \context, \seed) = \mtfun(norm(\carddist{c}(l, \context)), \carddist{q}(l, \context, \seed))
\end{equation*}

%% file: sections/experiments.tex
\vspace{10pt}
\section{Experimental Evaluation}
\label{sec:experiments}

\begin{figure*}[t!]
    \centering
    \begin{subfigure}[t]{0.49\textwidth}
      \includegraphics[width=\textwidth]{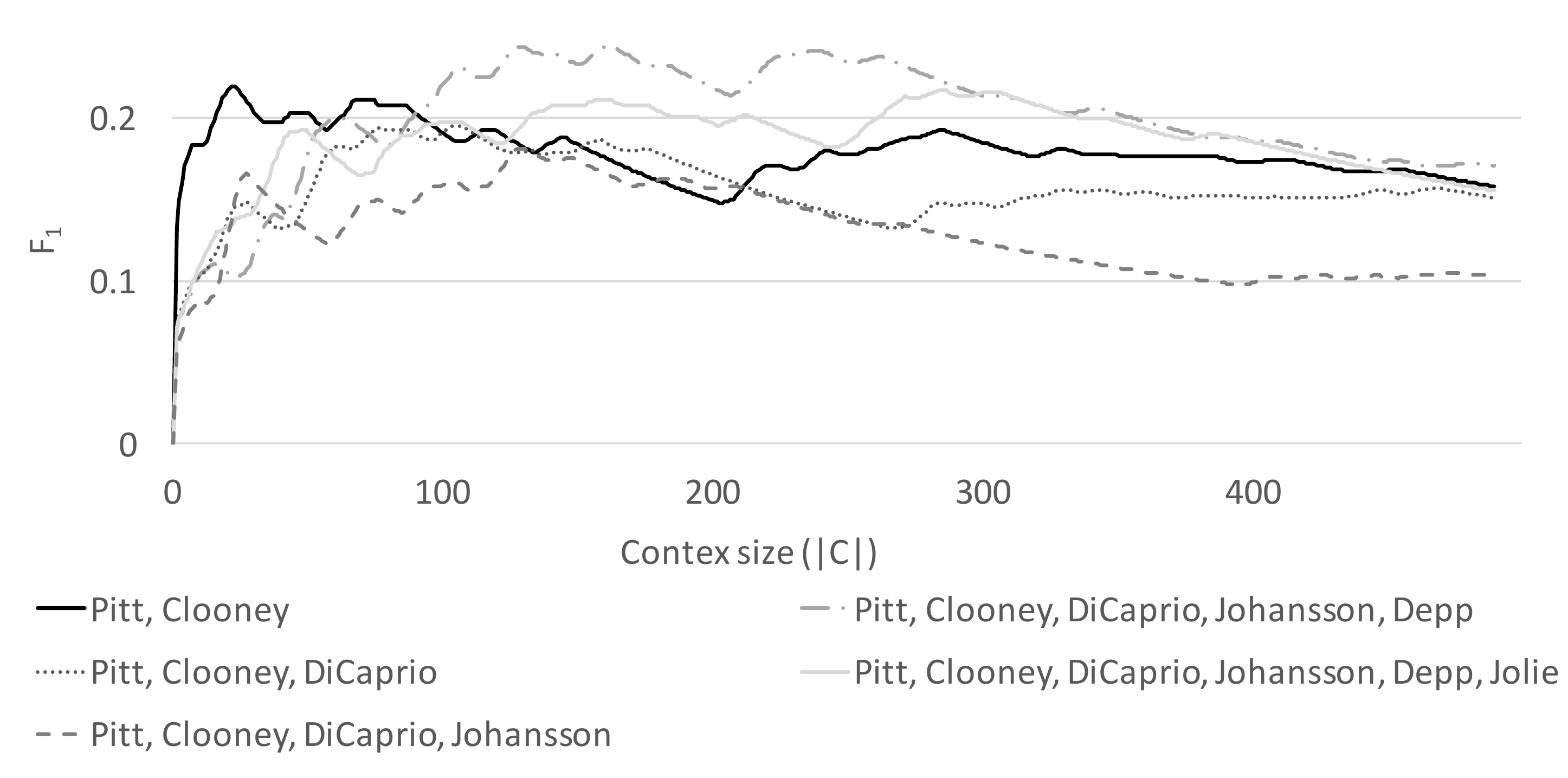}
      \caption{\ouralgorithm\ algorithm}
      \label{fig:crw_quality}
    \end{subfigure}
    \begin{subfigure}[t]{.49\textwidth}
        \includegraphics[width=\textwidth]{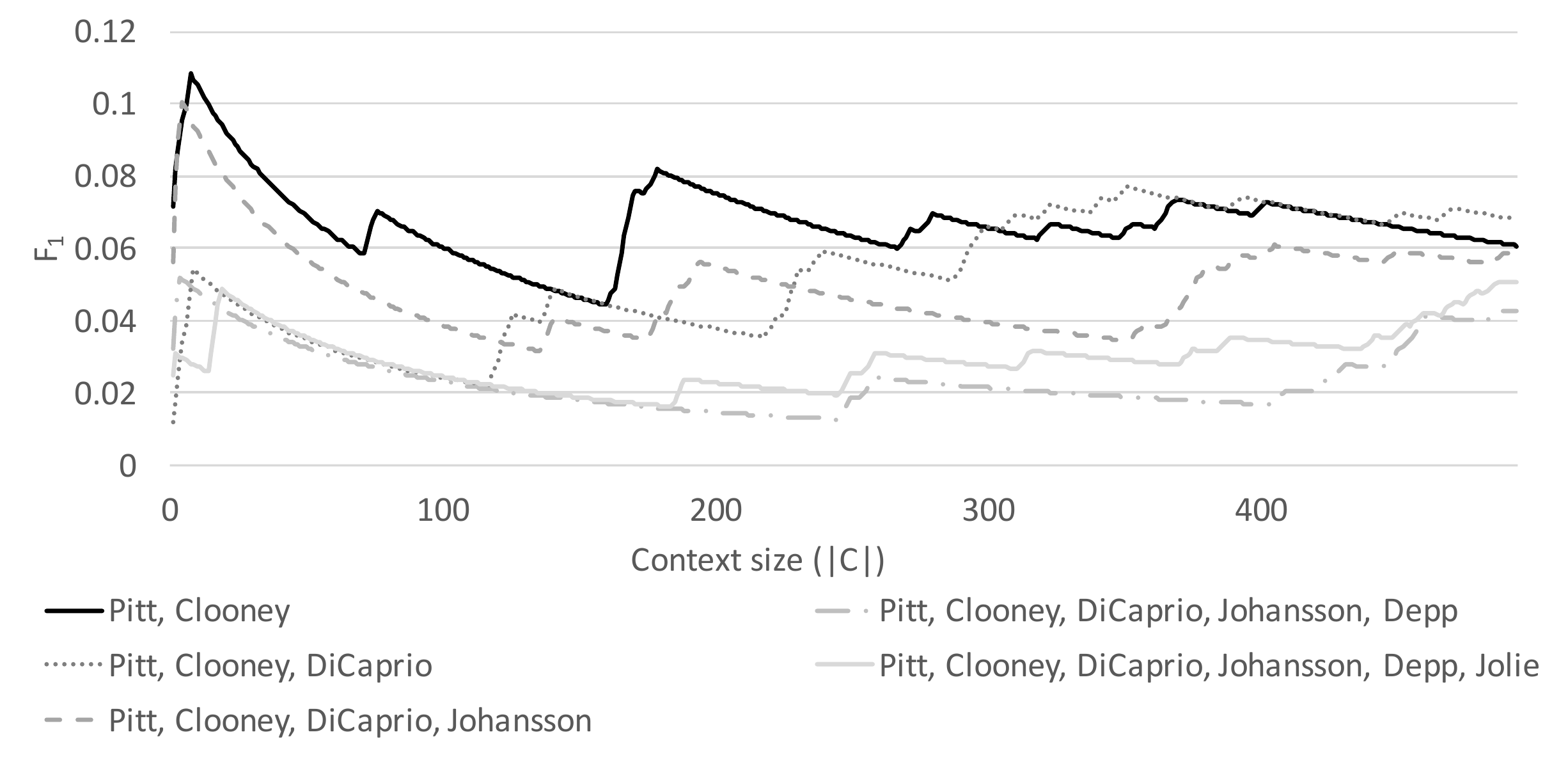}
      \caption{\rw\ algorithm}
      \label{fig:rw_quality}
    \end{subfigure}
    \caption{Quality ($F_1$) varying context size ($|\context|$) comparing \ouralgorithm\ and \rw\ with \actors\ domain in \yago\ dataset.} 
    \label{fig:rw_actor_quality}
\end{figure*}

In this section, we experimentally evaluate our approach on different datasets and show the impact of the parameters on the final results. 
Since there has been no other study on notable characteristics search so far, we have to generate a ground for evaluation. 
We do so by hiring users from a crowdsourcing platform and asking to manually provide context nodes as Wikipedia entities. 
We then mapped the entities to the corresponding nodes in one of the largest knowledge graphs available.

\spara{Datasets:}
We perform experiments on two datasets: \yago\ and \lmdb.

\noindent $\bullet$ \yago{} \cite{biega2013inside} is a large knowledge graph based on Wikipedia, Wordnet and Geonames. 
We downloaded YAGO 2.5\footnote{\url{http://resources.mpi-inf.mpg.de/yago-naga/yago2.5/yago2s_tsv.7z}} core facts in April 2016. 
It consists of 3.3M nodes and 27M edges, with 366K node types and 38 edge labels, including a type-hierarchy for node types.
As described in Section~\ref{sec:problem}, we represented each node attribute as an edge, having the attribute value as node label. 

\noindent $\bullet$ \lmdb{} is a knowledge graph for the movie domain, extracted from the Internet Movie Database (IMDB). 
We downloaded a snapshot of \lmdb\footnote{\url{https://datahub.io/dataset/linkedmdb}} in June 2016.
It consists of 739K nodes and 1.6M edges of 18 types.

\spara{Experimental Setup:}
We implemented our solution in Java 1.8, and ran the experiments on a machine with a quad-core Intel i5-4210U CPU 1.7 GHz and 12GB RAM. 
All the datasets are loaded into Apache Jena triple store to perform quick traversals on the graph without loading it into main memory. 
The impemented algorithms are the following.

\noindent $\bullet$ \rw: A baseline algorithm for context selection based on Personalized PageRank as described in Section~\ref{ssec:context}. 
Instead of the matrix multiplication we used the more scalable power iteration method. 
We set the number of iterations to 10 and the damping factor $c = 0.2$.

\noindent $\bullet$ \ouralgorithm: 
This is our algorithm described in Section~\ref{ssec:context}, that includes PathMining to mine the metapaths, the weighted random walk constrained to the metapaths found by PathMining, and the final score. 
We ran PathMining 1M times to retrieve the relevant metapaths. 

\noindent $\bullet$ \findnotable: 
This is our algorithm that incorporates \ouralgorithm\ and our method described in Section~\ref{ssec:distributions}. 
For the multinomial test, we used a statistic package written in R.

\spara{Summary of the experiments:}
We evaluate our algorithms effectiveness by comparing the retrieved context nodes to the ground truth obtained through our user survey.
Our context selection returns a better context compared to the baseline quicker. 
Moreover, our algorithm performs better as the query size increases. 
The returned notable characteristics indeed represent interesting undisclosed facts in the query nodes.

\begin{figure}[t!]
    \centering
    \includegraphics[width=.49\textwidth]{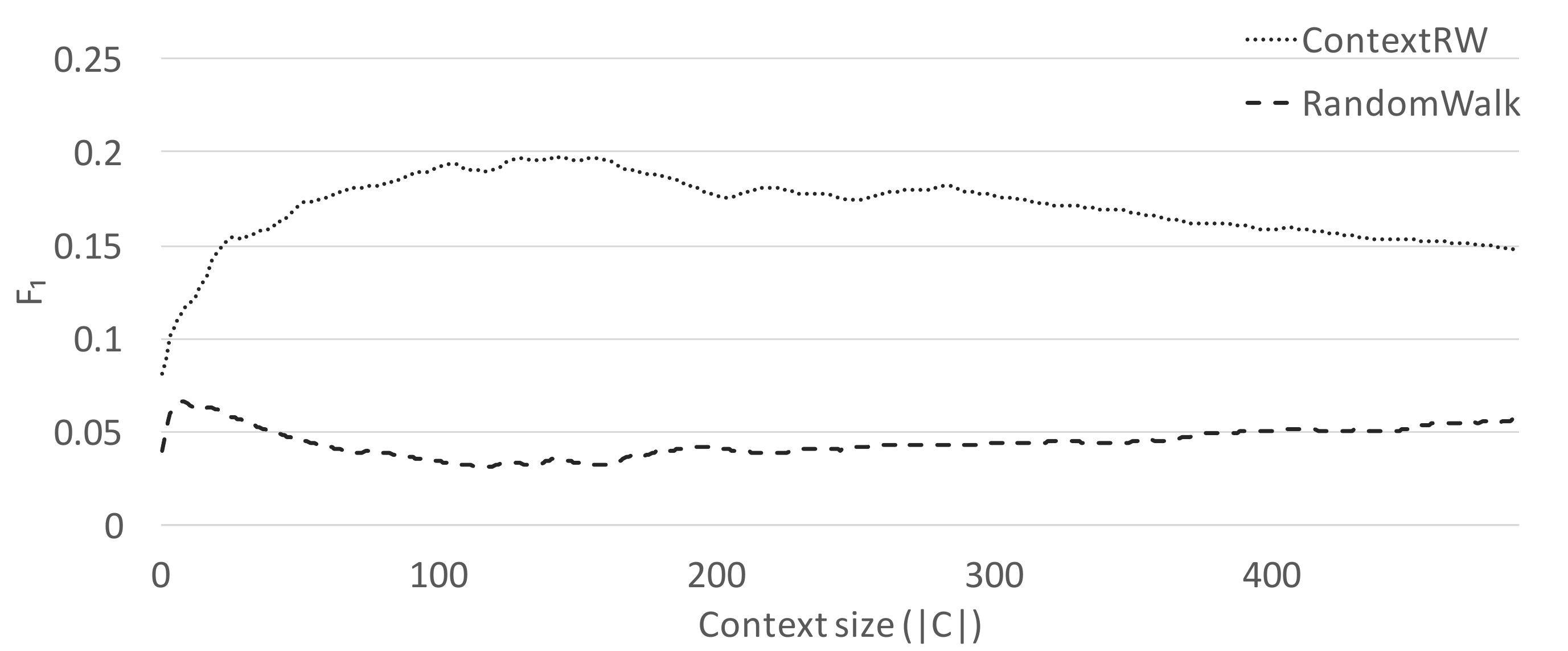}
    \caption{Average quality ($F_1$) vs context set size ($|\context|$) comparison in \yago\ dataset.}
    \label{fig:compare_contextsize}
\end{figure}

\begin{table}[!t]
  \centering
  \small
  \begin{tabular}{|c|c|c|}
    \hline
    \emph{\politicians} & \emph{\actors} & \emph{\moviepeople} \\
    \hline
   Angela Merkel  & Brad Pitt         & Steven Spielberg\\
   Barack Obama   & George Clooney    & Robert Downey Jr.\\
   Vladimir Putin & Leonardo DiCaprio & Hans Zimmer\\
   David Cameron  & Scarlett Johansson& Quentin Tarantino\\
   Fran\c{c}ois Hollande &Johnny Depp & Ellen Page\\
   Xi Jinping     & Angelina Jolie    & Celine Dion\\
   \hline
    \end{tabular}
  \caption{Entities in the three domains used in the evaluation.}
  \label{tab:scenarios}
\end{table}

\subsection{Evaluating Context selection}

We compare the effectiveness of \ouralgorithm\ with the baseline \rw\ within different topics. 
Unfortunately, to the best of our knowledge there was no existing ground for evaluation, i.e., finding context nodes given a query. 
This is crucial in our case, since the results with a single node can be dramatically different than those obtained with multiple query nodes. 
For instance, if the query only contains US presidents, we expect to find a context of US presidents.
On the contrary, if the query comprises both US and German politicians, we expect to find even politicians from other countries. 

Therefore, we generated the first ground for evaluation by crowdsource contexts for given query nodes.
We selected 15 query sets from three domains, namely \emph{\politicians}, \emph{\actors}, and \emph{\moviepeople}, to evaluate the algorithms. 
For each domain we manually determined 6 query nodes belonging to the domain, such as Angela Merkel and Barack Obama for \emph{\politicians}. 
The set of nodes (or entities) for each domain is shown in Table~\ref{tab:scenarios}.
We generated a ground truth with increasing query size by asking real users to provide a ranked list of entities related to those provided in the query. 
We hired 34 workers for each test set, asking them to provide 15 entities each.
For this experiment we used the CrowdFlower\footnote{\url{https://www.crowdflower.com/}} platform. 
This resulted in 510 entities for each of the 15 test sets (starting from 2 entities for each domain, adding one every time), with a total of 7'650 entities. 
After performing the manual labeling, we removed the entities mentioned only once, resulting in 36 to 76 entities for each query. 
Furthermore, we noted that for the politician scenario, our version of \yago\ misses some recent facts, e.g., Fran\c{c}ois Hollande is not mentioned as a head of state.
Hence, we manually substituted the name of the current head of state with the one found in \yago{}. 
This substitution does not substantially change the final result, since it preserves the structural properties in the graph, but allows us to evaluate the results with respect to the user knowledge. 

We evaluated the effectiveness of both \ouralgorithm\ and \rw\ in terms of $F_1$ score, which is defined as
\begin{equation*}
  F_1 = 2 * \frac{\text{precision} * \text{recall}}{\text{precision} + \text{recall}}
\end{equation*}

\begin{figure}[!t]
    \centering
    \includegraphics[width=0.49\textwidth]{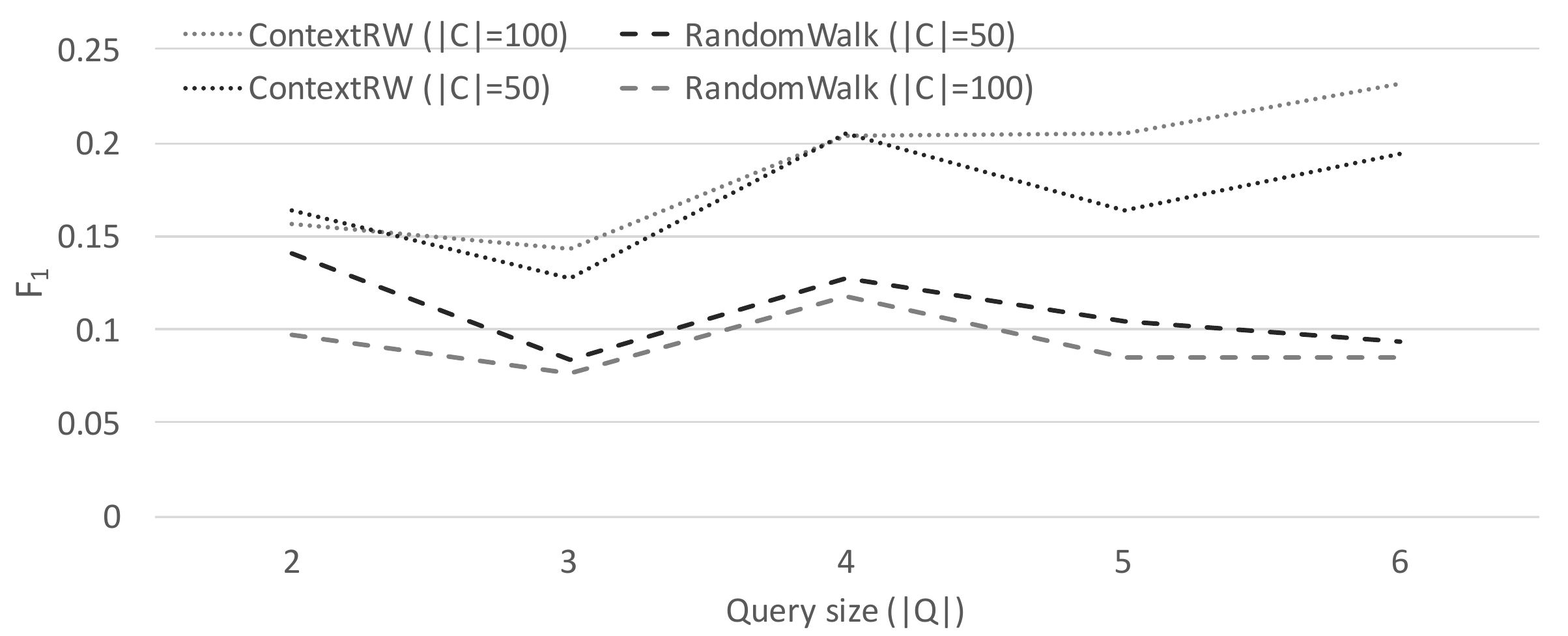}
    \caption{Average quality ($F_1$) vs query size ($|\seed|$) comparison in \yago\ dataset.}
    \label{fig:compare_seedsize}
\end{figure}

\begin{table}[t!]
  \centering
\begin{tabular}{c|c|c|c|}
\rowcolor{gray!50}$|\seed|$ & & $\max F_1$& $|\context|$ \\
  \hline
2 & \yago & 0.23 & 23 \\
  \cline{2-4}
 & \lmdb & 0.30 & 101 \\
  \hline
3 & \yago & 0.2 & 107 \\
  \cline{2-4}
 & \lmdb & 0.25 & 122 \\
  \hline
4 & \yago & 0.19 & 130 \\
  \cline{2-4}
 & \lmdb & 0.24 & 124 \\
  \hline
5 & \yago & 0.25 & 162 \\
  \cline{2-4}
 & \lmdb & 0.26 & 198 \\
  \hline
6 & \yago & 0.22 & 285 \\
  \cline{2-4}
 & \lmdb & 0.25 & 139 \\
  \hline
\end{tabular}
  \caption{Comparing the performance of \ouralgorithm{} on \yago{} and \lmdb{} in the \emph{\actors} domain.}
  \vspace{-10pt}
  \label{tab:yago_vs_linkedmdb}
\end{table}

\mpara{Context size ($|\context|$).} 
Context size $|\context|$ affects the quality of the results, since the more context nodes potentially the better recall but worse precision. 
While we report the quality in terms of $F_1$ score, we do not report time performance, since the number of context nodes generated is always the same for every run of the algorithms. 
Therefore, the results report the $F_1$ score at different cutoffs in the ranked context set. 
Figure~\ref{fig:rw_actor_quality} shows the $F_1$ score for varying the size of the context set ($|\context|$) for the queries in \emph{\actors} domain, using \yago\ dataset. 
We compare the performance of our \ouralgorithm\ (Fig.~\ref{fig:crw_quality}) with respect to the baseline \rw\ (Fig.~\ref{fig:rw_quality}). 
In all cases, \ouralgorithm\ performs 2 times better than the baseline, indicating that the metapath constrained random walk actually improves the overall quality. 
This is because many close neighbors of the query nodes are irrelevant considering the similarity notion between the query nodes, and this information is ignored by the simple \rw. 
After an initial increase in quality, we experience a non increasing trend when the context is bigger than 100 nodes.
This is motivated by an increasing recall as more context nodes are considered, but a drop in precision. 
We also note that the \rw\ algorithm shows a higher variance while \ouralgorithm\ is more stable.
The result is not surprising since \ouralgorithm\ includes only nodes within metapaths, while the baseline explores the space randomly, increasing the overall noise. 
In Figure~\ref{fig:compare_contextsize} we compare the average quality of the two algorithms using the entire query. 
The results show that \ouralgorithm\ is on average better than \rw\ in terms of quality, performing up to four times better for context size $|\context| = 100$. 

\begin{figure}[t!]
    \centering
    \includegraphics[width=0.48\textwidth]{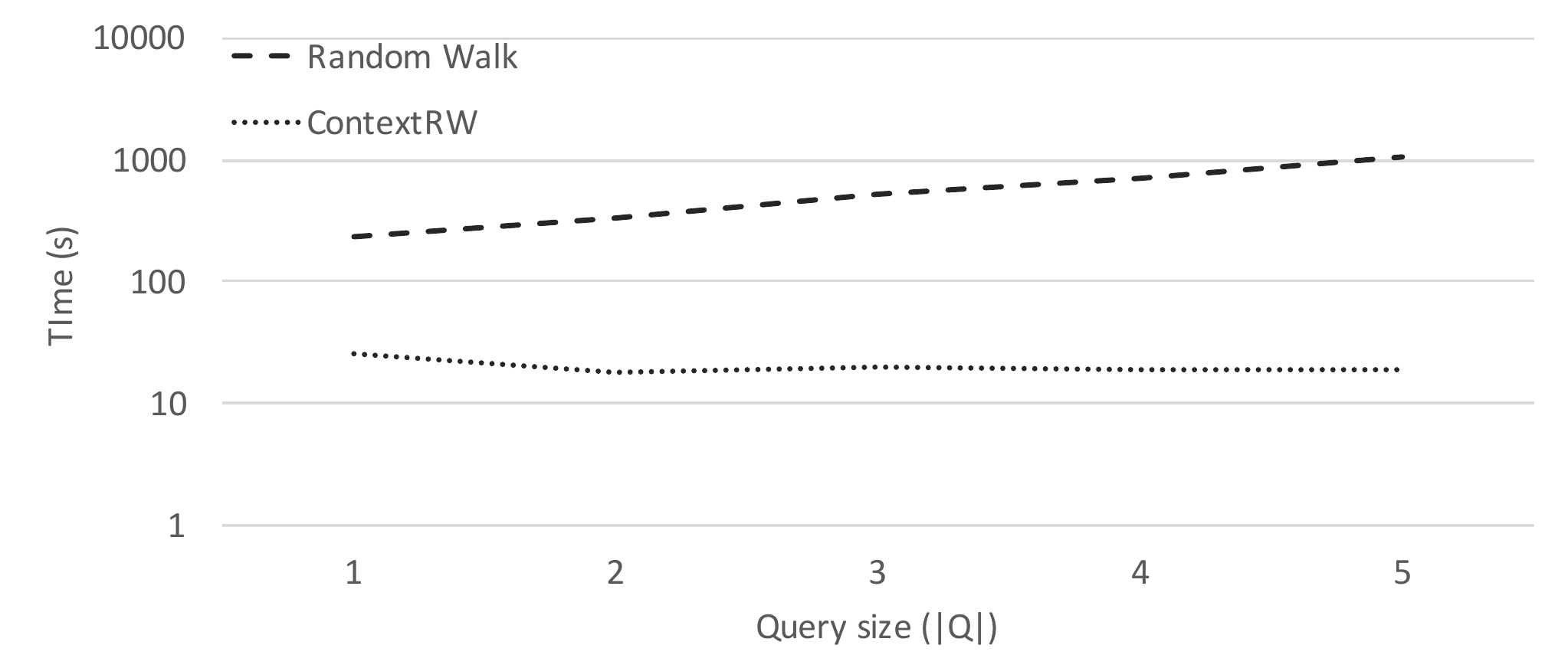}
    \caption{Average time (s) vs query size ($|\seed|$) comparison in \yago\ dataset.}
    \label{fig:timeVsSeed}
\end{figure}

\begin{figure}[t!]
    \centering
    \includegraphics[width=0.5\textwidth]{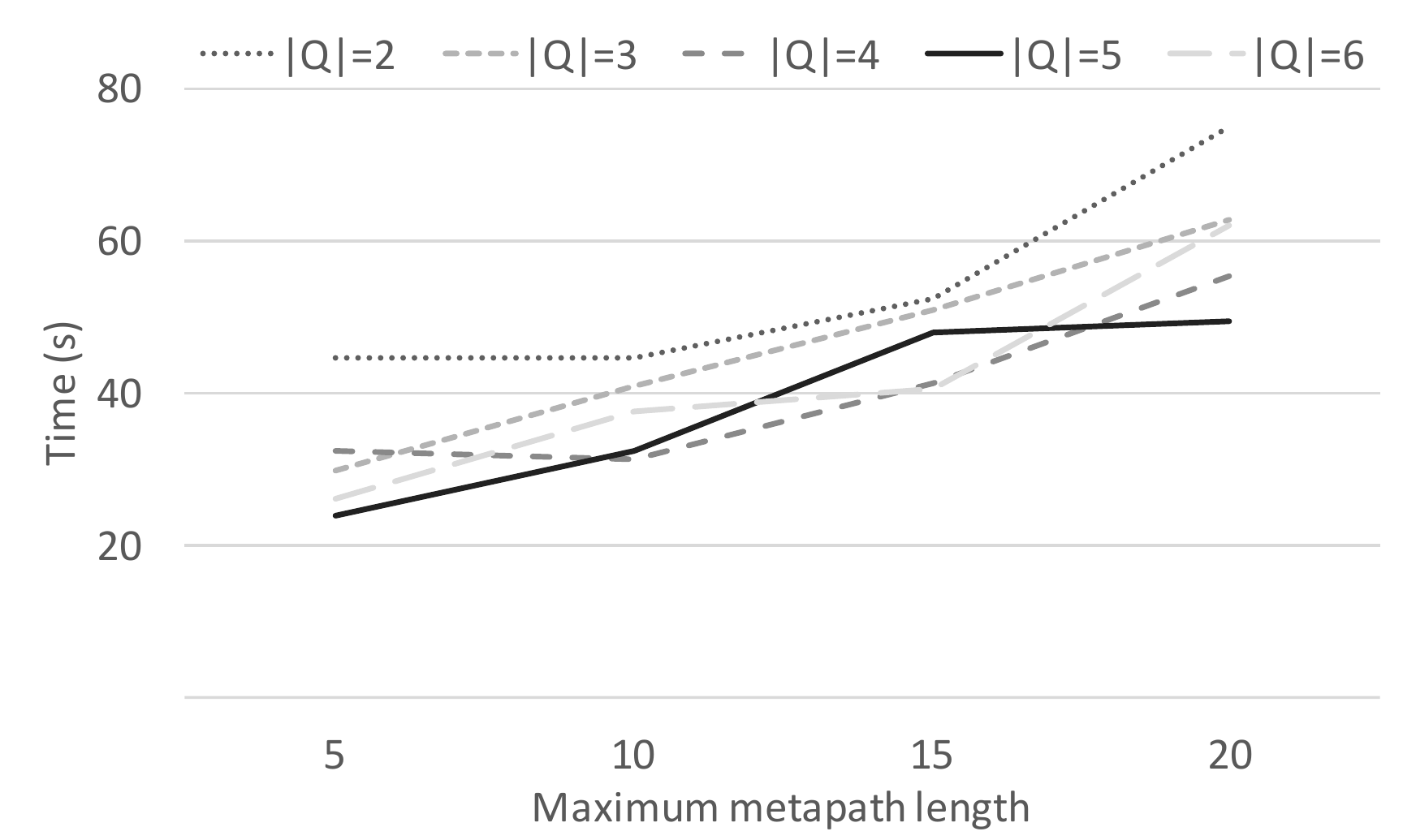}
    \caption{Time (s) vs metapath length with different query size ($|\seed|$)}
    \vspace{-10pt}
    \label{fig:pm_time_metapathlength}
\end{figure}

\mpara{Query size ($|\seed|$).} The query size $|\seed|$ affects both time and quality. 
We analyze the performance of the algorithms varying the query size. 
Figure \ref{fig:compare_seedsize} shows that \ouralgorithm{} improves in result quality when more query nodes are considered, supporting the claim that our method can capture semantic relationships between the nodes.
On the contrary \rw is not affected by the size of the query. 
This is a reasonable finding provided that \rw\ does not consider metapaths. 

Additionally, we compare the total runtime of each method varying the query size ($|\seed|$). 
Figure~\ref{fig:timeVsSeed} shows the time to compute the context for \ouralgorithm\ and the baseline \rw. 
We note that the \rw\ algorithm is on average up to two orders of magnitude slower than \ouralgorithm, for $|\seed| = 5$. 
Moreover, while \ouralgorithm\ is faster with larger queries, a random walk approach tends to become slower. 
This is an expected behavior in \ouralgorithm, since the chances to end the exploration in a query node is larger as the query size increases. 
Furthermore, we are able to return results in less than 20s.

Table~\ref{tab:yago_vs_linkedmdb} reports the maximum $F_1$ score at increasing $|\seed|$, comparing \yago\ and \lmdb\ datasets within the \emph{\actors} domain using the \ouralgorithm\ algorithm. 
While we could not evaluate for the \emph{\politicians} domain because the knowledge is not included in the \lmdb\ dataset, the results for \emph{\moviepeople} are mostly comparable and omitted for brevity. 
Unsurprisingly, \ouralgorithm\ performs better in \lmdb\ due to the specificity of the dataset.
However, the overall maximum increasing in $F_1$ is not larger than $0.7.$ 
This supports the claim that \ouralgorithm\ is able to capture domain specific knowledge even in more general datasets, exploiting the characteristics of the graph and the metapaths.  

\mpara{Number of paths ($|M|$).} The \ouralgorithm\ algorithm depends on the number of paths. 
Table~\ref{tab:f1_topk_pm} shows the $F_1$ score in relation to the context size and the number of paths. 
The number of paths does not affect the score; however, as shown in Figure~\ref{fig:pm_time_metapathlength} the time increases as the length of the metapaths (and also the number, not reported) increases. 
Therefore, a reasonable choice for the number of metapaths $|M|$ and maximum length is 5.

\begin{table}[t!]
  \centering
\begin{tabular}{c|c|c|c|c|}
  \multicolumn{1}{c}{} & \multicolumn{4}{c}{Number of paths ($|M|$)}\\
  \cline{2-5}
 $|\context|$ & 5 & 10 & 15 & 20 \\
 \hline
  50 &0.15 &0.16 &0.13& 0.15\\
100 &0.22 &0.21 &0.21& 0.21\\
150 &0.22 &0.23 &0.23& 0.23\\
200 &0.22 &0.22 &0.22& 0.22\\
  \hline
  \end{tabular}
  \caption{$F_1$ score as a function of the number of paths $|M|$ and the size of the context $|\context|$ for \ouralgorithm\ algorithm.}
  \label{tab:f1_topk_pm}
  \vspace{-10pt}
\end{table}

\subsection{Distribution Comparison}

We evaluate the performance of the \findnotable\ algorithm in terms of quality. 

\mpara{Metrics comparison.} We first evaluate the results comparing the characteristics found by \findnotable\ with those found by KL-divergence, and EMD that allow distribution comparison. 
We asked three human experts to provide a score to the characteristics of a small set of examples. 
We then aggregated the individual judgments and compared the ranking with the one obtained by the three methods. 
The minimum number of switches needed to transform one ranking to the other was used as a metric. 
We found that \findnotable\ required 2 changes, while KL-divergence and EMD required 4 and 5, respectively, supporting the choice of the multinomial test as a measure of quality.

\mpara{Test cases.} 
In practice \findnotable\ detects results that are more interesting than the one retrieved by the baseline \rw\ when equipped with the multinomial test. 
We refer to \rw\ with multinomial test as \rwtest.

One test case includes the scenario with the best $F_1$ score for the context construction, that has $\seed=\{George~Clooney,$ $Brad~Pitt,$ $Leonardo~DiCaprio,$ $Scarlett~Johansson,$ $Johnny$ $Depp\}$ as query.
We selected the top 100 nodes as the context.
The distribution comparison with multinomial test identified multiple edge labels, for which we provide a visual analysis of the findings.

\begin{figure}[t!]
    \centering
    \includegraphics[width=0.5\textwidth]{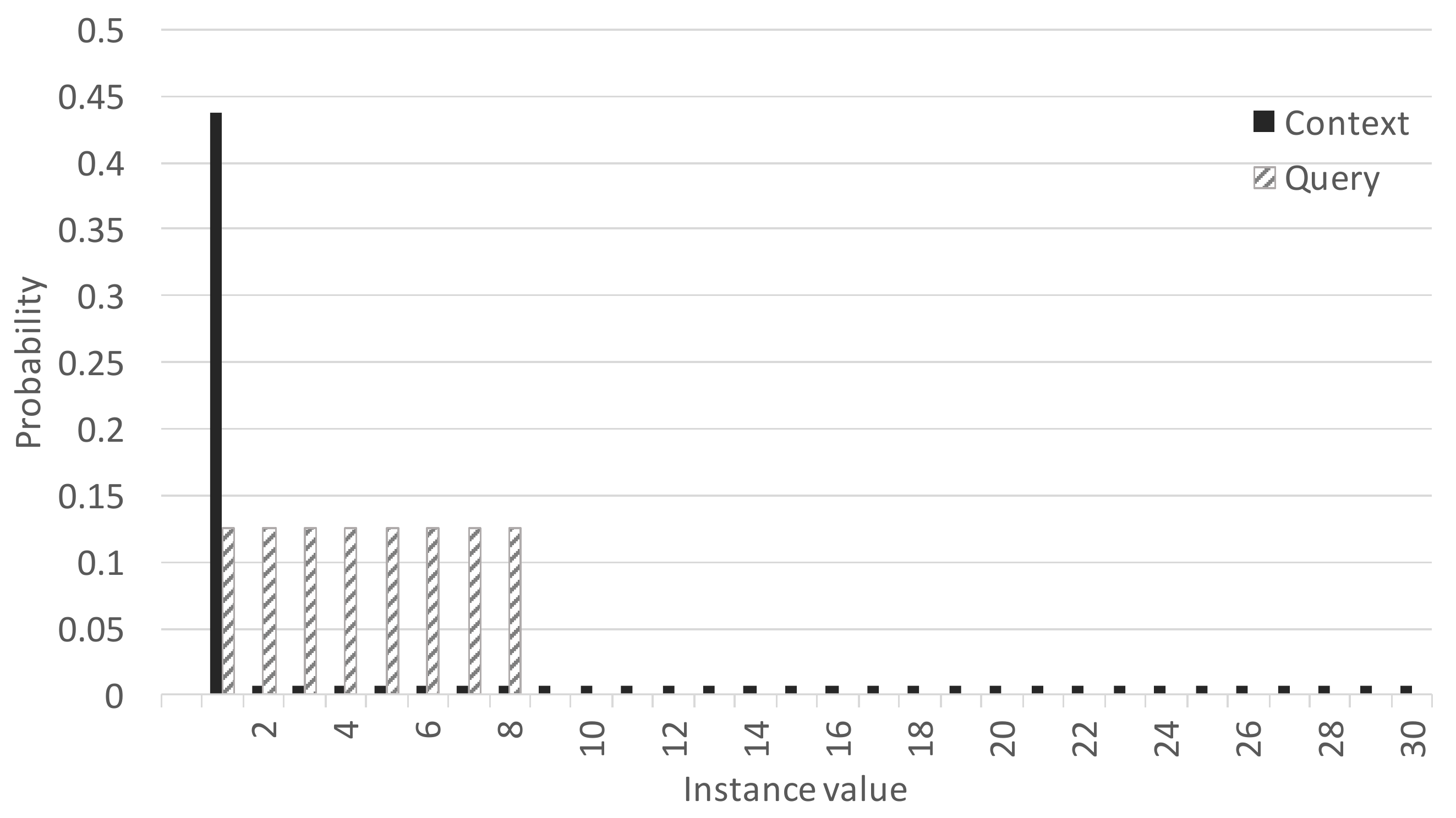}
    \caption{Distribution for the edge label \emph{created} with query \{Clooney, Pitt, DiCaprio, Johansson, Depp\} and $|C| = 100$. The first label is \emph{None}, indicating no matching edge found.}
    \label{fig:actors_created}
\end{figure}

\begin{figure}[t!]
    \centering
    \includegraphics[width=0.5\textwidth]{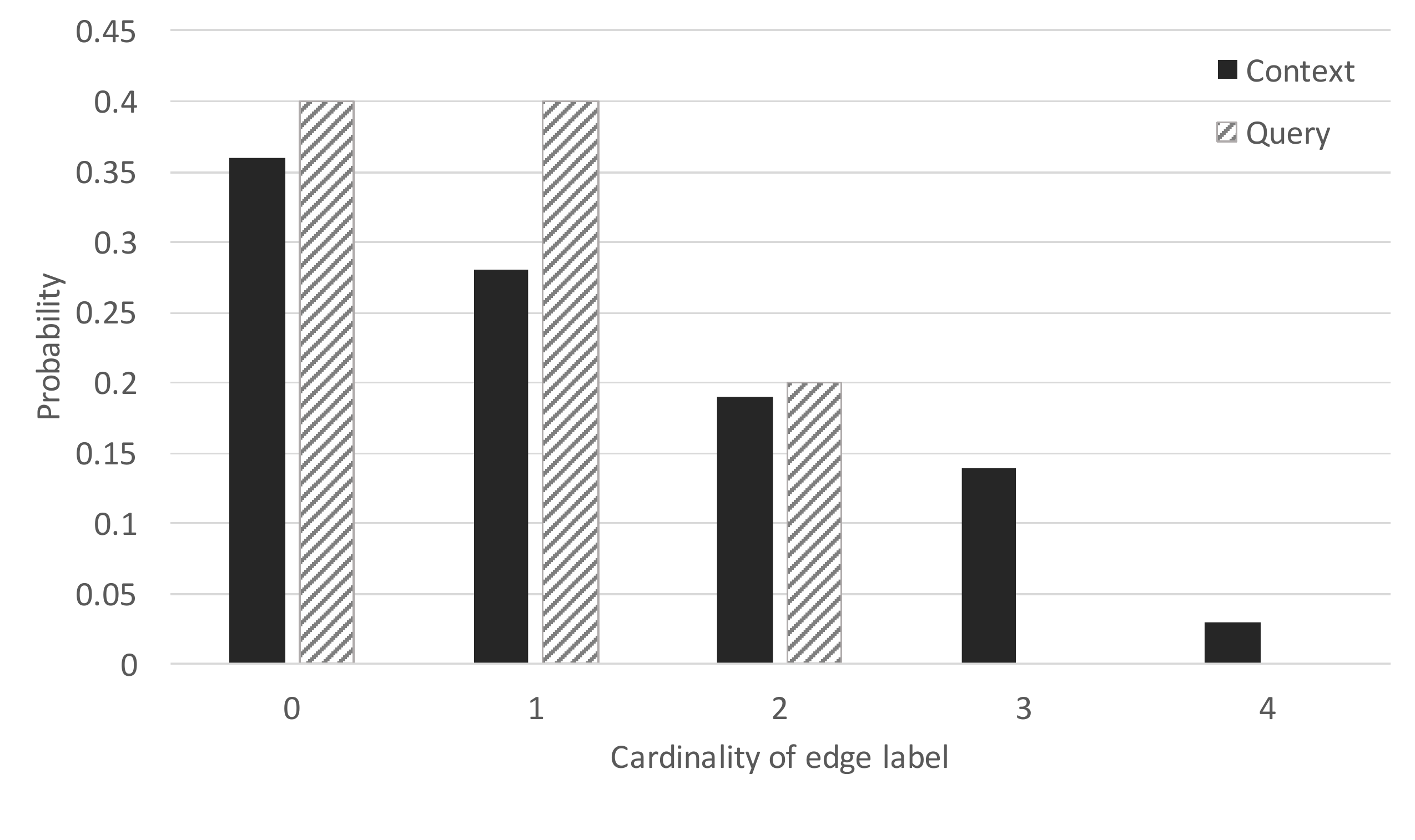}
    \caption{Distribution for the edge label \emph{hasWonPrize} with query \{Clooney, Pitt, DiCaprio, Johansson, Depp\} and $|C| = 100$.}
    \label{fig:actors_hasWonPrize}
    \vspace{-10pt}
\end{figure}

Figure~\ref{fig:actors_created} shows the instance distribution of the context for the \emph{created} edge label. 
The \emph{created} edge label is absent in 43\% cases (represented as \emph{None} instance), whereas all the other values are equally likely with 0.66\% chances. 
The query presents a different distribution, with one actor without \emph{created} labels and all the others with a different value. 
This marks a clear deviation from the context and is therefore identified as a notable characteristic by the multinomial test.
On the other hand, the \emph{hasWonPrize} edge, whose distributions are depicted in Figure~\ref{fig:actors_hasWonPrize}, is not marked as a notable characteristic.
Looking at the distributions for the context and the query, it is easy to see that they are quite similar.
The multinomial test cannot reject the null hypothesis of equality of the two distributions and therefore the \emph{hasWonPrize} edge-label is not notable.

In the second test case we use $\seed=\{Douglas~Adams,Terry$ $Pratchett\}$ as query and set the top 30 nodes as context. 
Our solution identified the edge \textit{influences} as a notable characteristic.
This is because the two authors in the query influenced an actor that was influenced by only 3 in total, and this result is definitely unexpected.
On the other hand, the edge \textit{created} was not found to be relevant.
All authors together created 834 works in total with only 3 of those being created by multiple authors.
As the query nodes also only created their own works and never collaborated, this is an expected result and thus not notable.

\begin{figure}[t!]
    \centering
    \includegraphics[width=0.5\textwidth]{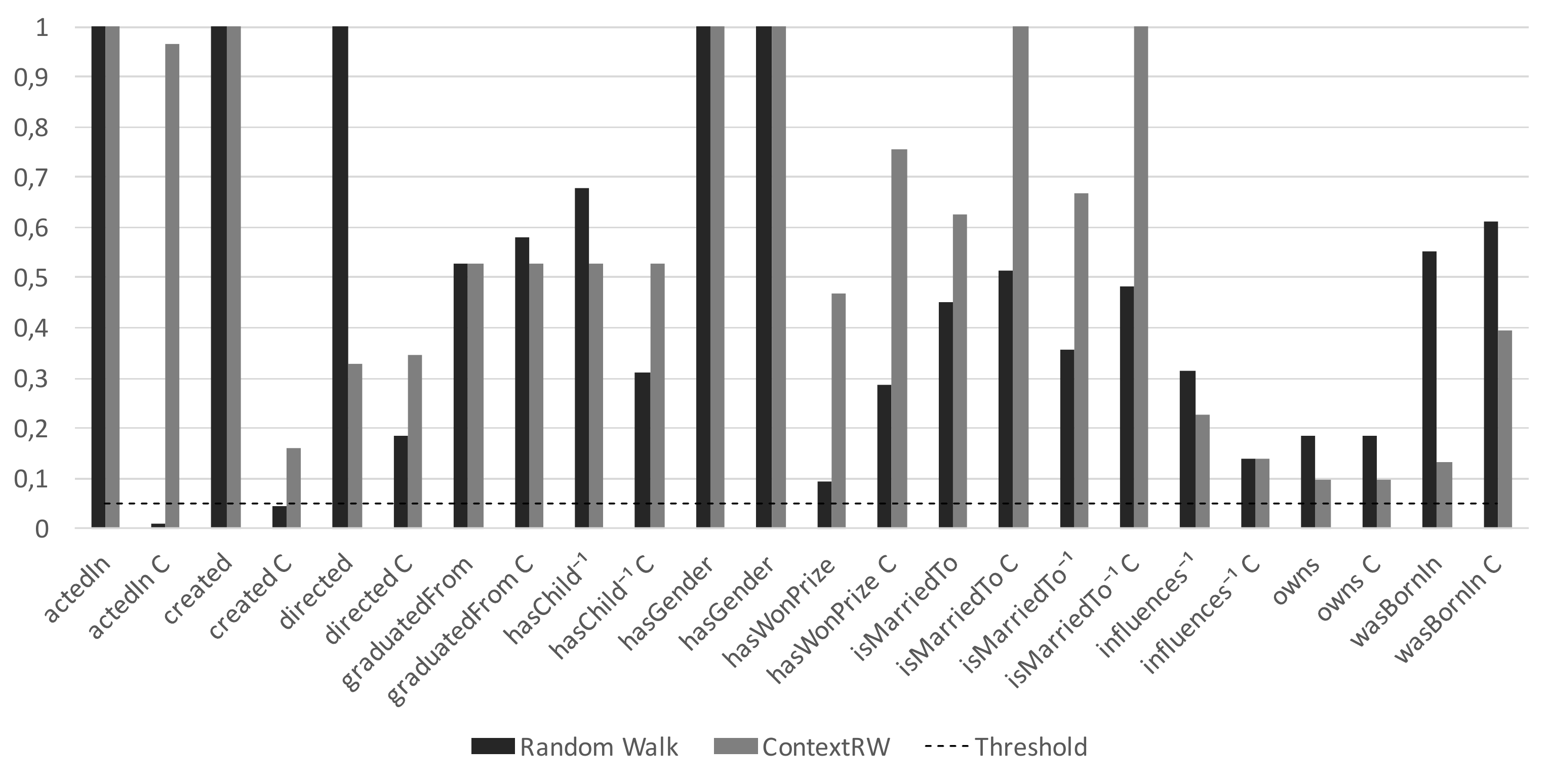}
    \caption{Comparison of significance probabilities for the \actors\ scenario with 5 query nodes. The ``C'' after the edge label denotes cardinality distributions.}
    \label{fig:actors_score_comp}
\end{figure}

\mpara{Algorithm comparison.} 
Figure \ref{fig:actors_score_comp} compares \findnotable\ with \rwtest, with query \{George Clooney, Brad Pitt, Leonardo DiCaprio, Scarlett Johansson and Johnny Depp\}.
All items above the threshold, depicted as a dashed line, are considered not interesting ($\discfun = 0$). 
The random walk selects mostly famous people in the movie business, therefore the \textit{actedIn} relation that connects actors with movies, is very rare in the context but common in the query, resulting in a score of 0.0086.
However, this is clearly not correct and, in fact, it is deemed as uninteresting by our \findnotable\ algorithm with a score of 0.96. 
Similarly, \textit{hasWonPrize} shows a significant difference between the two algorithms, as winning a prize is common for actors (75\%), but not so in the rather mixed random walk context (only 25\%).
The chart also shows that the significance level of the multinomial test can be used as a parameter to obtain the desired ``interestingness'' level.
Choosing 0.1 would include the \textit{owns} relationship as a notable characteristic, revealing that Brad Pitt is (according to the dataset) the only relevant actor to own a company (Plan B Entertainment).
This is specific for Brad Pitt, but not necessarily an interesting characteristic of the entire query, as it is reflected in the context.

%% file: sections/related.tex
\section{Related Work}
\label{sec:related}

Previous work on graphs mostly concerns the discovery of similar nodes or groups of individuals sharing common interests (graph clustering). 
In this section, we survey the most related works in these areas.

\mpara{Node comparison measures.}
Node comparison has a long history in graph analytics. 
Being able to compare pairs of nodes returning a similarity or distance score is a fundamental activity for clustering, ranking and classification. 
One of the earliest methods to compare nodes is graph edit distance~\cite{breiger1975algorithm} (GED), which is the minimum number of operations to transform a graph into another. 
Single nodes are compared in terms of the surrounding nodes and edges. 
Structural equivalence~\cite{lorrain1971structural} defines two nodes as similar if they have similar neighbors.
The first algorithm for structural equivalence is CONCUR~\cite{breiger1975algorithm}.
A similar approach is the one proposed by SimRank~\cite{jeh2002simrank}, which returns a self-similarity matrix between all the pairs of nodes in the graph. 
Random walk approaches, such as Personalized PageRank~\cite{chakrabarti2007dynamic} and HITS~\cite{kleinberg1999authoritative} can also be used to find nodes similar to the input nodes. 
Role discovery~\cite{henderson2012rolx} elaborates over the idea and, instead of returning a score they return multiple roles in terms of structural properties or graph global measures.
Node comparison measures can only return whether one node is similar or different from another but they cannot readily adapt to the discovery of notable characteristics, since the score provides no insight on the discovery process.
Additionally, these methods do not consider whether similarities or differences are meaningful with respect to a ``normal'' state other than total equivalence.

\mpara{Seed set expansion.}
Seed set expansion refers to methods that ask the user to provide an initial set of entities or structures and retrieve similar nodes. 
These methods, also known as example-based methods, can discover tuples of an unspecified result set in a relational database~\cite{dimitriadou2014explore,shen2014discovering}.
In graphs, the seed set can be composed of either structures (i.e., subgraphs) or nodes. 
Exemplar queries paradigm~\cite{mottin2014exemplar,mottin2016holistic} assumes that the user input is an example of the intended results.
Similarly, GQBE~\cite{jayaram2015querying} considers entity tuples to find similar other tuples in a knowledge graph.
These works are orthogonal to the discovery of notable characteristics, for they merely return answers similar to the input. 

Seed nodes are used to discover groups of nodes with similar characteristics~\cite{kloumann2014community,perozzi2014focused}. 
These seed-based clustering algorithms exploit the specificity of each node in the seed set to return ad-hoc communities. 
Likewise, seed-based approaches are used to discover dense regions in the graph~\cite{gionis2015bump,ruchansky2015minimum}.
Although these methods provide multiple groups of nodes they cannot properly explain the characteristics and the differences among them; in general, they do not directly compare the query nodes with the others.

\mpara{Relevant path summarization.}
Our problem reminisces the discovery of path templates (or metapaths) between nodes. 
A metapath is a sequence of node and edge labels that abstract connections between nodes. 
As such, they express connection patterns that have been shown to be effective in capturing non-trivial relationships and user preference patterns, improving the quality of recommendation results~\cite{lee2012pathrank,sun2011pathsim}.
Methods have been proposed to automatically discover metapaths from a given seed set~\cite{akoglu2013mining,meng2015discovering}.
Metapaths can express notable connections between seed nodes, but are insufficient for the given problem.
They cannot express the lack of an edge (e.g., Angela Merkel has no children), nor they detect characteristics related to instances (e.g., Angela Merkel is female while most leaders are male).
Discovery of metapaths also ignores difference-based characteristics, such as two people born in different places when the majority of similar people was born in the same place.
Most importantly, metapath discovery lacks the evaluation of notability.
Notability does not correlate to frequency per se: Being born in the same place is only notable, if most similar people are born in other places.

%% file: sections/conclusions.tex
\section{Conclusions}\label{sec:conclusions}
In this paper, we study the problem of notable characteristics search given a set of query nodes in a knowledge graph.
A notable characteristic is a special property in the query nodes that makes them different from their similars. 
Our problem is twofold: We first find a context set that represents the nodes similar to the query nodes; we then identify the notable characteristics with a novel probabilistic framework. 
We devise an algorithm for context selection based on random walk and metapath discovery and prove its effectiveness and efficiency with real data and user generated ground truth. 
In order to find the notable characteristics, we propose a probabilistic notion that first computes distributions for each edge label and subsequently performs a multinomial test to mark the characteristics that deviate from the expected behavior. 
We show different test cases to demonstrate the applicability and the effectiveness of our approach in real dataset. 

As future work we plan to expand the notion of notable characteristics to incorporate more complex patterns. 
We also intend to explore correlations between attributes as well as graph structures and incorporate results into the model.

%% file: paper.bbl
\begin{thebibliography}{10}

\bibitem{akoglu2013mining}
L.~Akoglu, D.~H. Chau, C.~Faloutsos, N.~Tatti, H.~Tong, J.~Vreeken, and
  L.~Tong.
\newblock Mining connection pathways for marked nodes in large graphs.
\newblock In {\em SDM}, pages 37--45, 2013.

\bibitem{biega2013inside}
J.~Biega, E.~Kuzey, and F.~M. Suchanek.
\newblock Inside yago2s: A transparent information extraction architecture.
\newblock In {\em WWW}, pages 325--328, 2013.

\bibitem{bordino2013machu_picchu}
I.~Bordino, G.~De~Francisci~Morales, I.~Weber, and F.~Bonchi.
\newblock From machu\_picchu to rafting the urubamba river: anticipating
  information needs via the entity-query graph.
\newblock In {\em WSDM}, pages 275--284. ACM, 2013.

\bibitem{breiger1975algorithm}
R.~L. Breiger, S.~A. Boorman, and P.~Arabie.
\newblock An algorithm for clustering relational data with applications to
  social network analysis and comparison with multidimensional scaling.
\newblock {\em Journal of mathematical psychology}, 12(3):328--383, 1975.

\bibitem{chakrabarti2007dynamic}
S.~Chakrabarti.
\newblock Dynamic personalized pagerank in entity-relation graphs.
\newblock In {\em WWW}, pages 571--580, 2007.

\bibitem{dimitriadou2014explore}
K.~Dimitriadou, O.~Papaemmanouil, and Y.~Diao.
\newblock Explore-by-example: An automatic query steering framework for
  interactive data exploration.
\newblock In {\em SIGMOD}, pages 517--528, 2014.

\bibitem{gionis2015bump}
A.~Gionis, M.~Mathioudakis, and A.~Ukkonen.
\newblock Bump hunting in the dark: Local discrepancy maximization on graphs.
\newblock In {\em ICDE}, pages 1155--1166, 2015.

\bibitem{henderson2012rolx}
K.~Henderson, B.~Gallagher, T.~Eliassi-Rad, H.~Tong, S.~Basu, L.~Akoglu,
  D.~Koutra, C.~Faloutsos, and L.~Li.
\newblock Rolx: structural role extraction \& mining in large graphs.
\newblock In {\em KDD}, pages 1231--1239, 2012.

\bibitem{hua2015short}
W.~Hua, Z.~Wang, H.~Wang, K.~Zheng, and X.~Zhou.
\newblock Short text understanding through lexical-semantic analysis.
\newblock In {\em ICDE}, pages 495--506, 2015.

\bibitem{jayaram2015querying}
N.~Jayaram, A.~Khan, C.~Li, X.~Yan, and R.~Elmasri.
\newblock Querying knowledge graphs by example entity tuples.
\newblock {\em TKDE}, 27(10):2797--2811, 2015.

\bibitem{jeh2002simrank}
G.~Jeh and J.~Widom.
\newblock Simrank: a measure of structural-context similarity.
\newblock In {\em KDD}, pages 538--543, 2002.

\bibitem{kasneci2009ming}
G.~Kasneci, S.~Elbassuoni, and G.~Weikum.
\newblock Ming: Mining informative entity relationship subgraphs.
\newblock In {\em CIKM}, pages 1653--1656, New York, NY, USA, 2009.

\bibitem{kleinberg1999authoritative}
J.~M. Kleinberg.
\newblock Authoritative sources in a hyperlinked environment.
\newblock {\em JACM}, 46(5):604--632, 1999.

\bibitem{kloumann2014community}
I.~M. Kloumann and J.~M. Kleinberg.
\newblock Community membership identification from small seed sets.
\newblock In {\em KDD}, pages 1366--1375, 2014.

\bibitem{kullback1951information}
S.~Kullback and R.~A. Leibler.
\newblock On information and sufficiency.
\newblock {\em The annals of mathematical statistics}, 22(1):79--86, 1951.

\bibitem{lee2015pathmining}
S.~Lee, S.~Lee, and B.-H. Park.
\newblock Pathmining: A path-based user profiling algorithm for heterogeneous
  graph-based recommender systems.
\newblock In {\em FLAIRS Conference}, pages 519--523, 2015.

\bibitem{lee2012pathrank}
S.~Lee, S.~Park, M.~Kahng, and S.-g. Lee.
\newblock Pathrank: a novel node ranking measure on a heterogeneous graph for
  recommender systems.
\newblock In {\em CIKM}, pages 1637--1641, 2012.

\bibitem{lissandrini2015unleashing}
M.~Lissandrini, D.~Mottin, T.~Palpanas, D.~Papadimitriou, and Y.~Velegrakis.
\newblock Unleashing the power of information graphs.
\newblock {\em SIGMOD Record}, 43(4):21--26, 2015.

\bibitem{lorrain1971structural}
F.~Lorrain and H.~C. White.
\newblock Structural equivalence of individuals in social networks.
\newblock {\em The Journal of mathematical sociology}, 1(1):49--80, 1971.

\bibitem{meng2015discovering}
C.~Meng, R.~Cheng, S.~Maniu, P.~Senellart, and W.~Zhang.
\newblock Discovering meta-paths in large heterogeneous information networks.
\newblock In {\em WWW}, pages 754--764, 2015.

\bibitem{mottin2014exemplar}
D.~Mottin, M.~Lissandrini, Y.~Velegrakis, and T.~Palpanas.
\newblock Exemplar queries: Give me an example of what you need.
\newblock {\em PVLDB}, 7(5):365--376, 2014.

\bibitem{mottin2016holistic}
D.~Mottin, A.~Marascu, S.~B. Roy, G.~Das, T.~Palpanas, and Y.~Velegrakis.
\newblock A holistic and principled approach for the empty-answer problem.
\newblock {\em VLDB J.}, pages 1--26, 2016.

\bibitem{perozzi2014focused}
B.~Perozzi, L.~Akoglu, P.~Iglesias~S{\'a}nchez, and E.~M{\"u}ller.
\newblock Focused clustering and outlier detection in large attributed graphs.
\newblock In {\em KDD}, pages 1346--1355, 2014.

\bibitem{pound2012interpreting}
J.~Pound, A.~K. Hudek, I.~F. Ilyas, and G.~Weddell.
\newblock Interpreting keyword queries over web knowledge bases.
\newblock In {\em CIKM}, pages 305--314, 2012.

\bibitem{ruchansky2015minimum}
N.~Ruchansky, F.~Bonchi, D.~Garc{\'\i}a-Soriano, F.~Gullo, and N.~Kourtellis.
\newblock The minimum wiener connector problem.
\newblock In {\em SIGMOD}, pages 1587--1602, 2015.

\bibitem{shen2014discovering}
Y.~Shen, K.~Chakrabarti, S.~Chaudhuri, B.~Ding, and L.~Novik.
\newblock Discovering queries based on example tuples.
\newblock In {\em SIGMOD}, pages 493--504, 2014.

\bibitem{sun2011pathsim}
Y.~Sun, J.~Han, X.~Yan, P.~S. Yu, and T.~Wu.
\newblock Pathsim: Meta path-based top-k similarity search in heterogeneous
  information networks.
\newblock {\em Proceedings of the VLDB Endowment}, 4(11):992--1003, 2011.

\bibitem{wang2004ontology}
X.~H. Wang, D.~Q. Zhang, T.~Gu, and H.~K. Pung.
\newblock Ontology based context modeling and reasoning using owl.
\newblock In {\em PerCom}, pages 18--22, 2004.

\end{thebibliography}
